\documentclass[final,5p,times,twocolumn]{elsarticle}

\usepackage{latexsym}
\usepackage{amssymb, amsmath, bm, physics}
\usepackage{subcaption}
\usepackage{mathptmx}
\usepackage{flushend}
\usepackage[colorlinks,citecolor=blue,urlcolor=blue,linkcolor=blue]{hyperref}
\usepackage{aas_macros}
\usepackage{orcidlink}

\biboptions{sort&compress}

\begin{document}

\begin{frontmatter}




\title{Resolution of challenging problems in quantum cosmology with electromagnetic radiation}


\author{S. Jalalzadeh \orcidlink{0000-0003-4854-2960}}
\address{Departamento de F\'{i}sica, Universidade Federal de Pernambuco, Recife-PE, 52171-900, Brazil}
\ead{shahram.jalalzadeh@ufpe.br}

\begin{abstract}
We investigate the quantum cosmology of a closed spatially homogeneous and isotropic Friedmann--Lema\^itre--Robertson--Walker (FLRW) minisuperspace model with electromagnetic radiation as matter content. We solve the corresponding Wheeler--DeWitt equation by utilizing Riemann’s zeta function regularization method. We demonstrate that the regularized vacuum energy of the electromagnetic field can overcome factor ordering, boundary conditions, and singularity problems.
\end{abstract}


\end{frontmatter}


\section{Introduction}
Diffeomorphism invariance raises stubborn obstacles in all quantum gravity approaches, with forms varying depending on the quantization method. One notably unregenerate issue is that the kinetic term in a diffeomorphism-invariant theory's Hamiltonian involves products of position and momentum variables. The issue of a proper factor ordering of operators is more than scholastic in mathematics. Various choices, for example, may result in different spectral features. This results in factor ordering ambiguities in a canonical approach to quantization, which gives us the Wheeler--DeWitt (WDW) equation because, as noted above, the kinetic term in this equation contains products of noncommuting conjugate operators. However, from a semiclassical perspective \cite{Bojowald:2014ija,Louko:1988zb}, it has been suggested that the factor ordering issue is not essential to the theory as a whole \cite{PhysRev.160.1113,PhysRevD.58.067301}.
Furthermore, alternative operator orderings \cite{Kontoleon:1998pw,Wiltshire:1998wb} produce different wavefunctions, and it has been suggested that the ordering issue is tied to a specific boundary condition \cite{Kontoleon:1998pw,Wiltshire:1998wb}. Unfortunately, there is no agreement on how to address this issue; thus, any recommendations provided are purely hypothetical.

The selection of an appropriate boundary condition for the universe's wavefunction has been a fundamental objective of quantum cosmology. Two `competitive' options are the no-boundary proposal \cite{PhysRevD.28.2960} and tunneling \cite{Vilenkin:1982de}. Despite their popularity, two alternative approaches to dealing with the occurrence of classical singularities have been used: the wavefunction \cite{PhysRev.160.1113} (or its derivative with respect to the scale factor \cite{1977TMP33.1076L}) vanishes at the classical singularity. It is important to remember that all of the aforementioned boundary conditions were developed on the spot, from a specific physical perspective. Most critically, these boundary conditions were not incorporated into the dynamical law. However, according to DeWitt \cite{PhysRev.160.1113}, `the constraints are everything,' implying that nothing more should be required.

Singularities, including cosmological ones, are the most exotic and enigmatic phenomena in General Relativity. A singularity in cosmology depicts a stage in the origin of the Universe where conditions are so severe that all currently known laws of physics fail. As a result, the unavoidable collapse of spacetime raises a fundamental issue that has yet to be resolved.
Quantum effects, in particular, have been proposed as a means of escaping the singularity in a classical collapse situation. DeWitt proposed \cite{PhysRev.160.1113} that the wavefunction of the Universe must vanish at the classical singularity of the appropriate classical cosmological model. This boundary condition, known as the DeWitt boundary condition, was the first criterion to avoid the classical singularity.
Furthermore, various authors proposed generalizing DeWitt's criteria to ensure conformal invariance; for example, \cite{Kiefer:2019bxk} and references therein. However, its fundamental relevance has long been a source of debate, and as Ref. \cite{PhysRevD.28.2402} has been shown, the DeWitt boundary condition has little to do with avoiding singularities. As recommended by Refs. \cite{PhysRevD.28.2402,PhysRevD.22.235,PhysRevD.8.3253}, to know the situation of quantum singularity, one can examine the expectation values of observables that classically vanish at the singularity of. Consequently, a quantum state $\psi$ is singular if and only if $\langle\psi|Q|\psi\rangle=0$ for every quantum observable, $ Q$, whose classical analog vanishes at the singularity. This test for quantum collapse is just as persuasive as any other and has the added benefit of being relatively easy to verify.

In this Letter, we provide new insight into the structure of quantum cosmology models, at least in the simplest form. In particular, we present a simple quantum cosmological model in the presence of electromagnetic radiation. We use Riemann's zeta function regularization method to show that the finite vacuum energy causes the model to pick up specific boundary conditions and ordering of operators. As a result, the expectation value of the scale factor is proportional to Planck's length, in which the constant of proportionality depends on the coefficient of regularized vacuum energy.

\section{Radiation in FLRW cosmology}
The Hamiltonian formalism of minimally coupled Maxwell's field cosmology in a closed homogeneous and isotropic universe is briefly reviewed in this section. This is necessary for the following section, in which we analyze the corresponding WDW equation.

Let us start with action functional for gravity and electromagnetic field with the standard form
\begin{multline}\label{1-1}
S=\frac{1}{16\pi G}\displaystyle\int_\mathcal{M}R\sqrt{-g}d^4x
\\-\frac{1}{4}\displaystyle\int_\mathcal{M}F_{\mu\nu}F^{\mu\nu}\sqrt{-g}d^4x+S_\text{GHY},
\end{multline}
where $S_\text{GHY}$ is the Gibbons--Hawking--York boundary term, $g$ is the determinant of spacetime manifold $\mathcal{M}$ which we assume is a closed, spatially homogeneous and isotropic with line element
\begin{eqnarray}\label{1-2}
ds^2=-N^2(t)dt^2+a^2(t)\Big(d\chi^2+\sin^2(\chi)d\Omega_{(2)}^2\Big).
\end{eqnarray}
The electromagnetic field tensor $F_{\mu\nu}$ is given in terms of the 4-vector potential $A_\mu$ as
$F_{\mu\nu}:=A_{\nu,\mu}-A_{\mu,\nu}$.
Using  the form of the metric (\ref{1-2}), the action functional (\ref{1-1}) can be rewritten as
\begin{equation}\label{1-4}
\begin{split}
   & S= \frac{3\pi}{4G}\displaystyle\int \Big\{-\frac{a\dot a^2}{N} +Na\Big\}dt+
       \frac{1}{2}\displaystyle\int \Big\{\frac{a}{N}\Big(h^{ij}\dot A_i\dot A_j
   \\ &+2A_0~\partial_t(^{(3)}\nabla^k A_{k})-A_0~^{(3)}\nabla^2A_0\Big)
         -\frac{N}{a}A^k\Big(2A_k\\&+~^{(3)}\nabla_k\,^{(3)}\nabla^iA_i-~^{(3)}\nabla^2A_k\Big)\Big\}\sqrt{h}dtd^3x,
\end{split}
\end{equation}
where the overdot denotes derivative with respect to the cosmic (comoving) time $t$, $h_{ij}$ are the components of the metric of the unit three-sphere $\mathbb S^3$,  
the spatial indices in (\ref{1-4}) are raised by the metric of unit 3-sphere and $^{(3)}\nabla$ denotes the induced spatial covariant derivative.  We cannot simply assume that the 4-potentials are only functions of cosmic time since radiation is made up of propagating waves in all directions that are not restricted (unless in the geometrical optics limit) to moving on spacetime geodesics \cite{1988GReGr201M}.  Thus $A_\mu=A_\mu(t,x^i)$. Because of the rotation group, $SO(4)$, the symmetry of $\mathbb S^3$, the time component of 4-potential $A_0$ can be expanded as a generalized Fourier expansion in terms of the scalar hyperspherical harmonics, which are eigenfunctions of the covariant Laplace operator on 3-sphere \cite{Lee}
\begin{eqnarray}\label{1-6}
^{(3)}\nabla^2 Y_{jlm}=-j(j+2)Y_{jlm},
\end{eqnarray}
where $0\leq l\leq j$, $-l\leq m\leq l$ with orthonormality conditions
\begin{eqnarray}\label{1-7}
\int \sqrt{h}d^3xY_{jlm}Y_{j'l'm'}=\delta_{jj'}\delta_{ll'}\delta_{mm'}.
\end{eqnarray}
Any scalar field, as well as $A_0$, on $\mathbb S^3$ can be written as
\begin{equation}\label{1-8}
A_0(t,x)=\sum_{j=0}^\infty\sum_{l=0}^j\sum_{m=-l}^{l}g_{jlm}(t)Y_{jlm}(x),    
\end{equation}
where $g_{jlm}(t)$ are functions of $t$.
In addition, the space components of the 4-potential, $A_i$, can be written as a expansion in terms of the vector hyperspherical harmonics, defined by the following three classes, denoted by $Y^{jlm}_{(B)n}$, $(B=0,1,2)$
\begin{equation}\label{1-9}
    \begin{split}
Y^{jlm}_{(0)i}&:=\frac{1}{\sqrt{j(j+2)}}~^{(3)}\nabla_iY^{jlm},
\\
Y^{jlm}_{(1)i}&:=\frac{1}{\sqrt{l(l+1)}}\varepsilon_i^{\,\,bc}~^{(3)}\nabla_bY^{jlm}\,^{(3)}\nabla_c\cos(\chi),
\\
Y^{jlm}_{(2)i}&:=\frac{1}{j+1}\varepsilon_i^{\,\,bc}~^{(3)}\nabla_bY^{jlm}_{(1)c}.
    \end{split}
\end{equation}
Then, the expansion of $A_k$ takes the form
\begin{eqnarray}\label{1-10}
A_k=\sum_{B=0}^2\sum_{j=j_\text{min}}^\infty\sum_{l=l_\text{min}}^j\sum_{m=-l}^l f^{jlm}_{(B)}(t)Y^{jlm}_{(B)k},
\end{eqnarray}
where $f^{jlm}_{(B)}(t)$ are only function of $t$ and $\varepsilon_{jlm}$ is the totally antisymmetric tensor volume element, $^{(3)}\nabla_j\varepsilon_{klm}=0$. Note that all hyperspherical vector harmonics vanish for $j=0$. Also, $Y^{jlm}_{(1),i}$ and $Y^{jlm}_{(2),i}$ are not well-defined for $l=0$. Therefore, these hyperspherical harmonics are defined only for $j\geqslant0$, and $l\geqslant0$ for $B=0$ and $l\geqslant1$ for $B=1,2$. From now on, to simplify the writing, we shall use $J$ to denote the set of indices $j$, $l$ and $m$, $J:=\{jlm\}$.  Inserting (\ref{1-8}) and (\ref{1-10}) into action (\ref{1-4}) and using the following properties of the vector harmonics \cite{Lee}
\begin{equation}\label{1-11}
\begin{split}
^{(3)}\nabla^2Y^J_{(0)i}&=(2-j(j+2))Y^J_{(0)i},\\
^{(3)}\nabla^2Y^J_{(1,2)i}&=(1-j(j+2))Y^J_{(1,2)i},\\
^{(3)}\nabla^iY^J_{(0)i}&=-\sqrt{j(j+2)}Y^J_{(0)i},\\
^{(3)}\nabla^iY^J_{(1,2)i}&=0,
\end{split}
\end{equation}
together with the orthonormality condition for vector hyperspherical harmonics
$\int h^{mn}Y^J_{(A)m}Y^{J'}_{(B)n}\sqrt{h}d^3x=\delta_{(A)(B)}\delta^{JJ'},$
the matter part of action (\ref{1-4}) simplifies to
\begin{multline}\label{1-13}
S_m=\displaystyle\int dt\Big\{\frac{a}{2N}\Big(\sum_{B=1}^2\sum_J\dot f^2_{(B)J}-2\sum_{BJ}g_{(B)J}\dot f_{(B)J}
\\
+\displaystyle\sum_{BJ}j(j+1)g^2_{(B)J}\Big)-\frac{N}{2a}\sum_{B=1}^2\sum_Jf^2_{(B)J}\Big\}.
\end{multline}
The corresponding momenta of $g_{(B)J}$ and $f_{(0)J}$ are obviously zero. As a result, they are Lagrange multiplies, which reflect the gauge freedom of the electromagnetic field in its canonical form. A simple way to eliminate gauge freedom is to set
\begin{eqnarray}\label{1-14}
f_{(0)J}=g_{(B)J}=0.
\end{eqnarray}
Eqs.(\ref{1-8}), (\ref{1-10}) and the last equation in (\ref{1-11}) show that the above conditions are equal to the following Coulomb-type (or radiation) gauge condition
\begin{eqnarray}\label{1-15}
A_0=0,\hspace{.5cm}^{(3)}\nabla^iA_i=0.
\end{eqnarray}
Thus, the action of electromagnetic field (\ref{1-13}) in terms of physical variables reduce to
\begin{multline}\label{1-16}
S_m=\displaystyle\sum_{B=1}^2\sum_{j=1}^\infty\sum_{l=1}^j\sum_{m=-l}^l\displaystyle\int dt\Big\{\frac{a}{ 2N}\dot f^2_{(B)J}
\\-\frac{ N}{2a}{(j+1)^2}f^2_{(B)J}\Big\},
\end{multline}
Eqs.(\ref{1-4}) and (\ref{1-16}) give us the total Lagrangian
\begin{multline}
    \label{1-17a}
    L=\frac{3\pi}{4G}\Big(-\frac{a\dot a^2}{N}+Na\Big)+
    \frac{1}{2}\sum_{B,j,l,m}\Big(\frac{a}{{N}}\dot f^2_{(B)J}-\\\frac{N}{a}(j+1)^2f^2_{(B)J}\Big).
\end{multline}
The field equations for the gauge field degrees of freedom will be
\begin{eqnarray}\label{1-18}
\frac{d}{dt}\left(\frac{a\dot f_{(B)J}}{ N}\right)+\frac{N(j+1)^2}{a}f_{(B)J}=0,
\end{eqnarray}
with solutions
\begin{equation}\label{1-19}
    f_{(B)J}=D_{(B)J}\sin\Big( (j+1) \eta+\theta\Big),
\end{equation}
where $D_{(B)J}$ and $\theta$ are the constants of integration, and the new time parameter $\eta$ is defined by $d\eta=\frac{N}{a}dt$.
In addition, the field equations for gravitational degrees of freedom, $\{a,N\}$, give us the Friedmann equations in the comoving frame ($N=1$)
\begin{equation}\label{1-20}
        \frac{\ddot a}{a}=-\frac{8\pi G}{3}\rho_0\left(\frac{a_0}{a}\right)^4,~~~
        H^2+\frac{1}{a^2}=\frac{8\pi G}{3}\rho_0\left(\frac{a_0}{a}\right)^4,
\end{equation}
where $H=\dot a/a$ is the Hubble parameter and
\begin{equation}\label{1-21}
    \rho_0:=\frac{1}{4\pi^2a_0^4}\sum_{(B)J}(j+1)^2D_{(B)J}^2,
\end{equation}
is the energy density of radiation at cosmic time $t_0$ where the scale factor was $a_0$. Apparently, the energy-momentum tensor of electromagnetic radiation, $T^{\mu\nu}$, seems to be in the form of a perfect fluid based on the above equations.  Let us look into this further. Using the Lagrangian density defined by (\ref{1-4}) in the standard definition of the energy-momentum tensor, we find the following components time-time, time-space and space-space components of $T^{\alpha\beta}$
\begin{equation}
   \begin{split}
     T^{00}&=\frac{1}{4\pi^2a^4N^2}\left(\frac{a^2}{N^2}h^{mn}\dot A_m\dot A_n+A_m(2A^m-~^{(3)}\nabla^2A_m)\right),\\
       T^{0i}&=0,\\
      T^{mn}&=\frac{1}{2\pi^2a^4N^2}\dot A^m\dot A^n-\frac{1}{2\pi^2a^4}\Big( 2A^mA^n-A^{(m}~^{(3)}\nabla^2A^{n)}-\\&A^k~^{(3)}\nabla^{(i}~^{(3)}\nabla^{j)}A_k\Big)+h^{mn}\Big(\frac{1}{4\pi^2a^4N^2}\dot A_k\dot A^k-\\&\frac{1}{4\pi^2a^4}(2A_kA^k-A^k~^{(3)}\nabla^2A_k)\Big).
   \end{split}
\end{equation}
By inserting (\ref{1-10}), (\ref{1-11}), (\ref{1-14}) and (\ref{1-19}) into the above equations, it is easy to verify that  $T_{ij}$ is not diagonal and consequently it is not share the same symmetries of the space-space components of the spacetime metric $g_{ij}$. Hence, $T_{\alpha\beta}$ do not have a perfect fluid form $T_{\alpha\beta}= (\rho+p)u_\alpha u_\beta+pg_{\alpha\beta}$ (where, $\rho$ and $p$ are energy density and pressure of the fluid respectively, and $u_\alpha$ is the 4-velocity). Thus, the geometrical symmetries in  Einstein tensor, $G_{\mu\nu}$ dose not accepted and imposed in the matter content. At classical level, to resolve this problem, we usually need to take integration over 3-sphere to obtain the average value for the energy-momentum tensor components. Then, the average values satisfy a perfect fluid with the equation of state $p=\frac{1}{3}\rho$, and the symmetries in Einstein tensor do accept with the perfect fluid \cite{1988GReGr201M}. On the other hand, the averaging has been done implicitly in obtaining the minisuperspace Lagrangian (\ref{1-17a}), and consequently, the right-hand side of the Friedmann equations (\ref{1-20}) realize the average properties of the radiation as a perfect fluid \cite{Man}. Also, at the semi-classical level, the Einstein field equations fulfill $G_{\alpha\beta}=8\pi G\langle 0|T_{\alpha\beta}|0\rangle$, where the expectation value of the energy-momentum tensor satisfies a perfect fluid form. An interesting discussion and generalization of the above ideas are presented in Refs.\cite{Bertolami:1990je,Bento:1992wy,Galtsov:1999bef,Moniz:1990hf,Bertolami:1991cf}. One question is: why do we not consider $A_i=A_i(t)$ and $A_0=g(t)$ from the start to ensure proper fitting with FLRW symmetries? If we choose a homogeneous and isotropic vector field, then the solution of field equations for the matter part will be $A_i(\eta)=D_i\sin(\sqrt{2}\eta)$, where $d\eta=\frac{N}{a}dt$, and $g(t)$ is an arbitrary function. This shows that the above choice leads to monochromatic radiation, which is in conflict with the cosmic microwave background (CMB) spectrum observations. Besides, to describe the black body radiation of CMB, we need a quantized radiation field coupled to gravity which that classically can be characterized by the action (\ref{1-16}). Let us add this comment that the matter part in the reduced ADM Lagrangian (\ref{1-17a}) carries the symmetries of the gravitational part as the result of the form of the original action functional (\ref{1-1}). For example, if we consider a Born--Infeld Lagrangian instead of the usual electromagnetic Lagrangian, then our reduction does not work anymore. More precisely, a general Born--Infeld gauge field Lagrangian is given by \cite{Galtsov:1999bef,Dyadichev:2001su,Moniz:2002rd,VargasMoniz:2003syv,VargasMoniz:2010upl}
\begin{equation}
    L_\text{BI}^\text{tr}\sim\beta^2\tr\left(1-\sqrt{1+\frac{1}{2\beta^2}F_{\alpha\beta}F^{\alpha\beta}-\frac{1}{16\beta^2}(F_{\alpha\beta}\tilde F^{\alpha\beta})^2} \right),
\end{equation}
where $\beta$ denotes the maximal field strength and $\tilde F_{\alpha\beta}$ is the dual of $F_{\alpha\beta}$. Since the gauge field $F_{\alpha\beta}$ is inside the root square, there is no way to obtain a reduced Lagrangian; consequently, these models are necessarily anisotropic (or inhomogeneous).

The total (ADM) Lagrangian (\ref{1-16}) leads us to the ADM Hamiltonian of the model
\begin{multline}
    \label{1-22}
    H_{ADM}=N\Big\{-\frac{G}{3\pi a}\Pi_a^2-\frac{3\pi}{4G}a+\\ 
    \frac{1}{2a}\sum_{(B)J}(\Pi_{(B)J}^2+(j+1)^2f_{(B)J}^2) \Big\}, 
\end{multline}
where $\Pi_a=-\frac{3\pi}{2G}\frac{a\dot a}{N}$ and $\Pi_{(B)J}=\frac{a}{N}\dot f_{(B)J}$ are the conjugate momenta of scale factor, $a$, and $f_{(B)J}$, respectively. The above ADM Hamiltonian leads us to the super-Hamiltonian constraint
\begin{multline}\label{1-23}
    \mathcal H=-\frac{G}{3\pi a}\Pi_a^2-\frac{3\pi}{4G}a+\\     \frac{1}{2a}\sum_{(B)J}\left(\Pi_{(B)J}^2+(j+1)^2f_{(B)J}^2\right)=0.
\end{multline}
 Let us define the
complex-valued functions
\begin{equation}
   \begin{split}\label{1-24}
   C_{(B)J}&:=\frac{1}{\sqrt{2(j+1)}}\left(i\Pi_{(B)j}+(j+1)f_{(B)J}\right),\\ 
   C^*_{(B)J}&:=\frac{1}{\sqrt{2(j+1)}}\left(-i\Pi_{(B)j}+(j+1)f_{(B)J}\right).
\end{split} 
\end{equation}
The set $S=\{C_{(B)J},C^*_{(B)J},1 \}$ is closed under the Poisson bracket $\{C^*_{(B)J},C_{(B')J'}\}=i\delta_{BB'}\delta_{JJ'}$, and every sufficiently differentiable function on the phase space of the matter sector can be expressed in terms of $S$. Now, the
super-Hamiltonian (\ref{1-23}) can be viewed as
\begin{multline}
    \label{1-25}
 H_{ADM}=N\Big\{-\frac{G}{3\pi a}\Pi_a^2-\frac{3\pi}{4G}a+\\ 
    \frac{1}{a}\sum_{(B)J}(j+1)C^*_{(B)J}C_{(B)J} \Big\}.
\end{multline}
The dynamics of these variables are given by
\begin{equation}
    \begin{split}
 &       C_{(B)J}(\eta)=C_{(B)J}(0)e^{-i(j+1)\eta},\\
 &       C_{(B)J}^*(\eta)=C^*_{(B)J}(0)e^{i(j+1)\eta}.
    \end{split}
\end{equation}
These solutions lead us to the mode-expansion of the gauge fields  (\ref{1-10})
\begin{equation}\label{1-27}
\begin{split}
A_i(\hat{x},\eta)=
\sum_{(B)J}\frac{1}{\sqrt{2(j+1)}}\Big\{C_{(B)J}Y_{(B)i}^Je^{-i(j+1)\eta}+\\ C^*_{(B)J}Y_{(B)i}^{J*}e^{i(j+1)\eta}\Big\}.
\end{split}
\end{equation}
The conjugate momenta of $A_i$ is $P^i=\delta S_m/\delta\dot A_i=\frac{ah^{ij}}{N}\dot A_j$ and the equal-time Poisson bracket is set to be $\{A_i(\eta,\hat{x}),P_j(\eta,\hat{ y})\}=\delta_{ij}(\hat{ x},\hat{{y}})$, where
\begin{eqnarray}\label{1-28}
\delta_{ij}(\hat{ x},\hat{{y}})=\sum_{(B)J}Y^{i*}_{(B)J}(\hat{{x}})Y^{j}_{(B)J}(\hat{{y}}),
\end{eqnarray}
is the delta function on $\mathbb S^3$.

\section{Quantum cosmology with radiation}

To obtain the WDW equation, we apply the quantization map 
\begin{equation}\label{quan}
\begin{split}
(a, \Pi_a)&\rightarrow (a,-i\frac{\partial}{\partial a}),\\ (f_{(B)J},\Pi_{(B)J})&\rightarrow (f_{(B)J},-i\frac{\partial}{\partial {f_{(B),J}}}).
\end{split}
\end{equation}
Moreover, we use 
the well-known Hartle--Hawking--Verlinde \cite{PhysRevD.28.2960,PhysRevD.33.3560,PhysRevD.37.888} factor ordering, defined by \footnote{There exist also two-parameter families of orderings in the literature  \cite{doi:10.1142/8540,Pedram:2008sj,Steigl:2005fk}. One can easily show that our general discussion in this letter can be extended to these kinds of orderings, and the results are the same as those we find here.  }
\begin{equation}
    \label{2-new1}
    \frac{1}{a}\Pi_a^2=a^{q-1}\Pi_aa^{-q}\Pi_a=-\frac{1}{a}\partial_a^2+\frac{q}{a^2}\partial_a,
\end{equation}
where $q$ is the factor ordering parameter.
Therefore, by applying the quantization map (\ref{quan}) to the Hamiltonian constraint (\ref{1-23}), we obtain the following WDW equation
\begin{multline}
    \label{2-29}
    \Bigg\{\frac{1}{3\pi a m_P^2}\left(-\partial_a^2+\frac{q}{a}\partial_a\right)+\frac{3\pi m_P^2}{4}a-
    \frac{1}{2a}\sum_{(B)J}\Big(-\partial_{(B)J}^2\\+(j+1)^2f_{(B)J}^2\Big)\Bigg\}\Psi(a,f_{(B)J})=0,
\end{multline}
where $m_P=1/\sqrt{G}$ is the Planck mass.
The matter part of the super-Hamiltonian,
\begin{equation}\label{2-31}
    \mathcal H_m=\frac{1}{2a}\sum_{(B)J}\left(-\partial_{(B)J}^2+(j+1)^2f_{(B)J}^2\right),
\end{equation}
in the above WDW equation, represents the contribution of infinite number of harmonic oscillators. The eigenfrequencies of the
electromagnetic field are $\omega_j=(j+1)/a$. If for the matter part of the super-Hamiltonian, instead of $(f_{(B)J},\Pi_{(B)J})$ we use the creation, $C^\dagger_{(B)J}$, and annihilation, $C^\dagger_{(B)J}$, operators, the quantum counterparts of the holomorphic functions (\ref{1-24}), in which 
\begin{eqnarray}\label{2-32}
    \left[C_{(B)J},C^\dagger_{(B')J'}\right]=i\delta_{BB'}\delta_{JJ'},
\end{eqnarray}
we can write (\ref{2-31}) as
\begin{multline}\label{2-33}
    \mathcal H_m=\frac{1}{a}\sum_{(B)J}(j+1)\left(N_{(B)J}+\frac{1}{2}\right)=\\\frac{1}{a}\displaystyle\sum_{B=1}^2\sum_{j=1}^\infty\sum_{l=1}^j\sum_{m=-l}^l(j+1)\left(N_{(B)jlm}+\frac{1}{2}\right)
\end{multline}
where $N_{(B)J}:=C^\dagger_{(B)J}C_{(B)J}$ are number operators.
According to the above Hamiltonian, the vacuum energy of the electromagnetic radiation is 
\begin{equation}\label{2-34}
    \mathcal H_m^\text{vacuum}=\frac{1}{a}\sum_{j=1}^\infty j(j^2-1).
\end{equation}
 This sum
is divergent because all the vacuum modes give contribution to the zero-point
energy.
We can regularize the divergent sum in the above expression of the vacuum energy by Riemann's zeta function regularization method \cite{Elizalde:1994gf} by defining
\begin{equation}\label{2-36}
\begin{split}
    \mathcal H_m^\text{re.vacuum}&=\frac{1}{a}\left(\sum_{j=1}^\infty j^{-3s}-\sum_{j=1}^\infty j^{-s} \right)\\
    &=\frac{1}{a}\Big(\zeta(3s)-\zeta(s)\Big),
    \end{split}
\end{equation}
where $s$ is a complex parameter with $Re(s)>1$ condition, and $\zeta(s)$ is the Riemann's zeta function. Using the expression \cite{Elizalde:1994gf}
\begin{equation}\label{2-37}
\lim_{\varepsilon\rightarrow0}\zeta(\varepsilon-m)= (-1)^m\frac{B_{m+1}}{m+1},~~~~~m\in\mathbb N,
\end{equation}
where $B_m$'s are Bernoulli numbers, we find 
\begin{eqnarray}
    \label{2-38}
    \mathcal H_m^\text{re.vacuum}=\frac{1}{a}\left(-\frac{B_4}{4}+\frac{B_2}{2} \right)=\frac{\lambda}{a},
\end{eqnarray}
where $\lambda=11/120$. Note that this result agrees with the result obtained by L.H. Ford in Ref. \cite{Ford:1976fn}.
Inserting this renormalized vacuum energy into the WDW equation (\ref{2-29}) simplifies it into
\begin{equation}\label{2-39}
    \left\{-\partial_a^2+\frac{q}{a}\partial_a+\frac{9\pi^2m_P^4}{4}a^2 \right\}\psi(a)=3\pi m_P^2\lambda\psi(a).
\end{equation}
 The operator in the left-hand side of this WDW equation is defined on a
dense domain $C^\infty(0,+\infty)$. Therefore, is not an essentially
self-adjoint operator. 
The necessary and sufficient condition for it to be an Hermitian operator is
\begin{equation}\label{2-40}
    {\psi(a)}\Big|_{a=0}={\gamma}\frac{d}{da}\psi(a)\Big|_{a=0},
\end{equation}
where $\gamma$ is a real parameter.
Changing the minisuperspace variable $a$, into $\xi=\frac{3\pi m_P^2}{2}a^2$ and rewriting the wavefunction as $\psi=\exp(-\frac{\xi}{2})g(\xi)$ in Eq.(\ref{2-39}) gives us
\begin{equation}
    \label{new2-2}
    \xi\frac{d^2g}{d\xi^2}+\left( \frac{1}{2}(1-q)-\xi\right)\frac{dg}{d\xi}-\left( \frac{1-q}{4}-\frac{\lambda}{2}\right)g=0.
\end{equation}
This equation is the well known Kummer’s differential equation \cite{Abramowitz}, whose regular solution
at the Big-Bang singularity, $\xi=0$, is the confluent hypergeometric function 
\begin{equation}\label{g11}
g(\xi)=~_1F_1\left(\frac{1-q-2\lambda}{4};\frac{1-q}{2};\xi\right).
\end{equation} 
In order for the wavefunction to be square-integrable, the hypergeometric series
\begin{equation}\label{newww}
    _1F_1(\alpha;\beta;\xi)=\sum_k^\infty\frac{\alpha^{(k)}\xi^k}{\beta^{(k)}k!},~~\alpha^{(k)}:=\alpha(\alpha+1)...(\alpha+k-1),
\end{equation}
must terminate. This requirement is satisfied if there exists some non-negative integers $n$ such that $\alpha=-n$, or equivalently in (\ref{g11})
\begin{eqnarray}
    \label{new2-3}
    \frac{1-q-2\lambda}{4}=-n,~~~~n=0,1,2,...~.
\end{eqnarray}
Furthermore, in this case, we have a representation of the confluent hypergeometric functions in terms of generalized Laguerre polynomials $\{L^{(\beta)}_n(\xi) \}_{n=0}^{\infty}$, given by \cite{Abramowitz}
\begin{equation}
    \label{new123}
   \binom{n+\beta}{n}~_1F_1(-n;\beta+1;\xi)=L^{(\beta)}_n(\xi).
\end{equation}
Therefore, the solution of the WDW equation (\ref{2-39}) is
\begin{equation}\label{2-41}
    \psi(a)=\mathcal N \exp\left\{-\frac{3\pi }{4l_P^2}a^2\right\}L^{(-\frac{1}{2}-\frac{q}{2})}_n\left(\frac{3\pi }{2l_P^2}a^2\right),
    \end{equation}
where $\mathcal N$ is a normalization constant and $l_P=1/m_P$ is the Planck length. A wavefunction (belonging the Hilbert space of physical states) is square-integrable if
\begin{eqnarray}
    \label{2-42}
 \int_0^\infty a^{-q}|\psi(a)|^2da<\infty.  
\end{eqnarray}
Inserting the explicit form of the wavefunction (\ref{2-41}) into the left-hand side of the above inequality, we find
\begin{multline}
    \label{new1122}
     \int_0^\infty a^{-q}|\psi(a)|^2da=\\\frac{\mathcal N^2}{2}\left(\frac{2}{3\pi m_P^2} \right)^{\frac{(1-q)}{2}}\int_0^\infty \xi^{-\frac{(1+q)}{2}}e^{-\xi}\left(L_n^{(-\frac{1}{2}-\frac{q}{2})}(\xi) \right)^2d\xi,
\end{multline}
where $\xi=\frac{3\pi m_P^2}{2}a^2$.
Regarding the orthogonality relation of generalized Laguerre polynomials
\begin{equation}
\begin{split}
 \frac{1}{\Gamma(1+\beta)}\int_0^\infty \xi^\beta e^{-x}L_m^{(\beta)}(\xi)L_n^{(\beta)}(\xi)d\xi=\binom{n+\beta}{n}\delta_{mn},  \\
  \text{if}~~\beta>-1,
 \end{split}
\end{equation}
the wavefunction is orthonormal if $-(1+q)/2>-1$. Concerning this circumstance on the values of the ordering parameter, $q$, and condition (\ref{new2-3}),
 we find that $n$ has to satisfy the inequality $n<\lambda/2=11/240$. Thus, the only acceptable value of $n$ is $n=0$, or equivalently, the value of the factor ordering parameter is 
\begin{equation}
    \label{2-43}
 q=1-2\lambda=\frac{49}{60}. 
\end{equation}
Therefore, the permitted normalized wavefunction is
\begin{equation}
    \label{2-44}
    \psi(a)=\sqrt{\frac{2}{\Gamma(\lambda)}}\left(\frac{3\pi m_P^2}{2}\right)^\frac{\lambda}{2}\exp\left\{-\frac{3\pi }{4l_P^2}a^2\right\}.
\end{equation}
Now, it is easy to see that the allowed boundary condition in (\ref{2-40}) is given by $\gamma=\infty$ (Neumann boundary condition), or
\begin{equation}
    \frac{d}{da}\psi(a)\Big|_{a=0}=0.
\end{equation}
In addition, the expectation value of the scale factor is
\begin{equation}
    \label{2-45}
    \langle a\rangle=\int_0^\infty a^{1-q}|\psi(a)|^2da=\frac{\Gamma(\lambda+\frac{1}{2})}{\Gamma(\lambda)}\sqrt{\frac{2}{3\pi}}l_P.
\end{equation}
This demonstrates that the quantum singularity has been removed. Besides, the wavefunction's normalization constant is controlled by the regularization parameter $\lambda$.
Generally, depending on our regularization method, it scales as $\frac{\Gamma(\lambda+\frac{1}{2})}{\Gamma(\lambda)}\simeq \sqrt{\lambda}$ for big $\lambda$ values. The scale factor's expectation value will be far from Planck length if more substantial $\lambda$ values have been used. In addition, according to Eq.(\ref{new2-3}), there is more freedom to set the ordering parameter $q$ in this circumstance. However, the scale factor expectation value does not vanish for any choice of $q$.


\section{Conclusions}
We show that the regularized vacuum energy of the electromagnetic field uniquely defines the state of the Wheeler--DeWitt equation in a closed homogeneous and isotropic universe. The only requirement we need to impose is that vacuum energy is finite, regardless of the regularization method used, resulting in a considerable reduction in the available wavefunctions. Furthermore, the expectation value of the scale factor is proportional to the Planck length.

Theoretical cosmologists typically employ a phenomenological description of matter's degree of freedom in the action (\ref{1-16}). For example, a perfect fluid \cite{1977TMP33.1076L,Alvarenga:2001nm,2002PhRvD65f3519B,Vakili:2010rf,Fathi:2016lws,2016IJMPD2530009J,Husain:2011tm,Husain:2019nym,Jalalzadeh:2014jka,Majumder:2011ad,Maeda:2015fna,Lawrie:2011eq,Demaerel:2019cao,Jalalzadeh:2014jea}, a Chaplygin gas \cite{Bouhmadi-Lopez:2011tfh,Bouhmadi-Lopez:2009ggt,Bouhmadi-Lopez:2004dni,Pedram:2007ud}, an anti-Chaplygin gas \cite{Kamenshchik:2013naa} or other phenomenological matters have been used. Because of the quantum nature of quantum cosmology, fundamental fields should theoretically characterize the matter content. In fact, quantum cosmologists defend such matter contents by stating that accurate general solutions are impossible to find in the presence of fundamental forces, the Hilbert space structure is ambiguous, and retrieving the concept of a semi-classical time is problematic. From the outset, it is evident that employing perfect fluid is essentially semi-classical. If we use these kinds of phenomenological matters in our model investigated here, the parameter $\lambda$ introduced in Eq.(\ref{2-38}), which is a free parameter, will represent the physical parameter of the matter field. For example, if we use dust or radiation as a perfect fluid, then $\lambda$ will realize the total mass of the dust and total entropy of the radiation, respectively \cite{doi:10.1142/8540,Jalalzadeh:2014jka,Jalalzadeh:2014jea}. As a consequence, Eq.(\ref{new2-3}) gives the mass or entropy, or other relevant parameters of the phenomenological matter fields in terms of quantum number $n$ and factor ordering parameter $q$ \cite{2016IJMPD2530009J}. The ordering parameter, $q$, is then arbitrarily chosen with some physical intuition.
Furthermore, the value of $\gamma$ in the boundary condition (\ref{2-40}) will be an arbitrary constant with a length dimension, which will be a new fundamental physical constant. To prevent creating a new fundamental constant, it must be set to zero or infinity. 

Regularization of the vacuum energy, on the other hand, provides us (depending on the regularization method) a unique and definite value of $\lambda$ by selecting $F_{\mu\nu}$ as a fundamental field.
Now, Eq.(\ref{new2-3}) gives us the operator ordering parameter, as well as the fixed value of quantum number $n$. Consequently, all these lead us to a specific wavefunction in the Hilbert space, in which the expectation value  of the scale factor is proportional to the Planck length. Instead of an electromagnetic field, if we consider a massless conformally coupled scalar field, we can show that $\lambda=1/240$. Thus the whole argument is also valid for such a scalar field.

Applying the zeta regularization method to massive vector fields with $So(3)$ global symmetries developed in Ref.\cite{Bertolami:1994jn} is also intriguing.
 In this case, in the vacuum energy (\ref{2-34}), the ground state oscillators' frequency and their mass will depend on the mass, say $\mu$, of the gauge field as well as the scale factor, $a$, simultaneously. Employing the zeta function regularization method (or other) developed in \cite{Elizalde:1994gf}, one can likewise obtain the contribution of vacuum energy of the corresponding gauge fields on the WDW equation. In these cases, the mass term's presence breaks the theory's conformal invariance, and the expectation value of the vacuum energy will be an infinite sum of polynomials of $\mu a$. Thus, our conclusions on the boundary conditions and the factor ordering cannot be extended to this case. 
  We are aware that our results are obtained within a straightforward cosmological model. Nevertheless, we think they are intriguing and provide motivation for subsequent research works. Possible extensions to test the zeta function regularization may include supersymmetric quantum cosmology with vector fields \cite{Moniz:1997qy,Moniz:1997zz},  the far more rich and elegant SU(2) ansatz in \cite{Bertolami:1991cf}, quantum cosmology with fermionic matter field \cite{DEath:1986lxx}, or full infinite-dimensional superspace \cite{Halliwell:1984eu}.

\section*{Acknowledgements}
The author would like to thank the anonymous reviewer for his/her insightful suggestions and careful reading of the manuscript.

\section*{Declaration of competing interest}
The author declares that he has no known competing financial interests or personal relationships that could have appeared to influence the work reported in this paper.

\bibliographystyle{elsarticle-num}
\bibliography{BH}

\end{document}